\documentclass[runningheads,envcountsame]{llncs}

\usepackage{amssymb,amsmath,mathtools,empheq,fancybox}
\newsavebox{\mybox}

\usepackage{paralist}
\usepackage{url}
\usepackage{caption}

\usepackage{textcomp,listings}

\iftrue 
\makeatletter
\let\old@lstKV@SwitchCases\lstKV@SwitchCases
\def\lstKV@SwitchCases#1#2#3{}
\makeatother
\usepackage{lstlinebgrd}
\makeatletter
\let\lstKV@SwitchCases\old@lstKV@SwitchCases

\lst@Key{numbers}{none}{%
    \def\lst@PlaceNumber{\lst@linebgrd}%
    \lstKV@SwitchCases{#1}%
    {none:\\%
     left:\def\lst@PlaceNumber{\llap{\normalfont
                \lst@numberstyle{\thelstnumber}\kern\lst@numbersep}\lst@linebgrd}\\%
     right:\def\lst@PlaceNumber{\rlap{\normalfont
                \kern\linewidth \kern\lst@numbersep
                \lst@numberstyle{\thelstnumber}}\lst@linebgrd}%
    }{\PackageError{Listings}{Numbers #1 unknown}\@ehc}}
\makeatother
\else
\usepackage{lstlinebgrd}
\fi

\usepackage{makecell}
\usepackage{longtable}
\usepackage[title]{appendix}
\usepackage{soul} 

\usepackage{xcolor}

\definecolor{color0}{RGB}{0, 0, 0} 
\definecolor{color1}{RGB}{186, 033, 033}
\definecolor{color2}{RGB}{000, 128, 000}
\definecolor{color3}{RGB}{0, 128, 255}
\definecolor{color4}{RGB}{170, 034, 255}
\def\truncdiv#1#2{((#1-(#2-1)/2)/#2)}
\def\moduloop#1#2{(#1-\truncdiv{#1}{#2}*#2)}
\def\modulo#1#2{\number\numexpr\moduloop{#1}{#2}\relax}

\makeatletter

\newcount\netParensCount@clisp

\lst@CCPutMacro
\lst@ProcessOther{`(}{{%
		\ifnum\lst@mode=\lst@Pmode\relax%
		\rainbow@clisp{(}%
		\global\advance\netParensCount@clisp by \@ne%
		\else
		(%
		\fi
}}%
\lst@ProcessOther{`)}{{%
		\ifnum\lst@mode=\lst@Pmode\relax%
		\global\advance\netParensCount@clisp by \m@ne%
		\rainbow@clisp{)}%
		\else
		)%
		\fi
}}%
\@empty\z@\@empty

\newcommand\rainbow@clisp[1]{%
	\ifcase\modulo\netParensCount@clisp 5\relax%
	\textcolor{color0}{#1}%
	\or
	\textcolor{color1}{#1}%
	\or
	\textcolor{color2}{#1}%
	\or
	\textcolor{color3}{#1}%
	\else
	\textcolor{color4}{#1}%
	\fi
}

\lst@AddToHook{PreInit}{%
	\global\netParensCount@clisp 0\relax%
}

\makeatother      
\definecolor{listinggray}{gray}{0.9}
\definecolor{lbcolor}{rgb}{0.9,0.9,0.9}
\definecolor{burgundy}{rgb}{0.5, 0.0, 0.13}

\lstdefinestyle{myCPP}{
	backgroundcolor=\color{lbcolor},
	tabsize=4,    
	language=[GNU]C++,
	basicstyle=\scriptsize,
	upquote=true,
	aboveskip={1.5\baselineskip},
	columns=fixed,
	showstringspaces=false,
	extendedchars=false,
	prebreak = \raisebox{0ex}[0ex][0ex]{\ensuremath{\hookleftarrow}},
	frame=single,
	numbers=left,
	showtabs=false,
	showspaces=false,
	showstringspaces=false,
	identifierstyle=\ttfamily,
	keywordstyle=\color[rgb]{0,0,1},
	commentstyle=\color[rgb]{0.026,0.112,0.095},
	stringstyle=\color[rgb]{0.627,0.126,0.941},
	numberstyle=\color[rgb]{0.205, 0.142, 0.73},
	morekeywords={NULL}
}

\lstdefinestyle{myJava}{
	backgroundcolor=\color{lbcolor},
	tabsize=4,    
	language=[GNU]C++,
	basicstyle=\scriptsize,
	upquote=true,
	aboveskip={1.5\baselineskip},
	columns=fixed,
	showstringspaces=false,
	extendedchars=false,
	prebreak = \raisebox{0ex}[0ex][0ex]{\ensuremath{\hookleftarrow}},
	frame=single,
	numbers=left,
	showtabs=false,
	showspaces=false,
	showstringspaces=false,
	identifierstyle=\ttfamily,
	keywordstyle=\color[rgb]{0,0,1},
	commentstyle=\color[rgb]{0.026,0.112,0.095},
	stringstyle=\color[rgb]{0.627,0.126,0.941},
	numberstyle=\color[rgb]{0.205, 0.142, 0.73},
}

\lstdefinelanguage{heapProlog}{
	backgroundcolor=\color{lbcolor},
	tabsize=4,    
	basicstyle=\scriptsize,
	upquote=true,
	aboveskip={1.5\baselineskip},
	columns=fixed,
	showstringspaces=false,
	extendedchars=false,
	escapeinside     = {(*}{*)},
	prebreak = \raisebox{0ex}[0ex][0ex]{\ensuremath{\hookleftarrow}},
	frame=single,
	numbers=left,
	showtabs=false,
	showspaces=false,
	showstringspaces=false,
	identifierstyle=\ttfamily,
	keywordstyle=\color[rgb]{0,0,1},
	commentstyle=\color[rgb]{0.026,0.112,0.095},
	stringstyle=\color[rgb]{0.627,0.126,0.941},
	numberstyle=\color[rgb]{0.205, 0.142, 0.73},
	alsoletter = {:,-},
	morekeywords={:-, emptyHeap, allocate, write, read, valid, true, false},
}

\lstdefinelanguage{SMT-LIB} {
	alsoletter       = {-, [=], >, <},
	morecomment      = [l]{;},
	morekeywords     = {define-fun, declare-const,define-sort, declare-heap,check-sat,ite,
		Bool,
		bvadd, bvsub, bvneg, bvsgt, bvsle,
		assert, and, or, not, let, =, =>,
	    declare-datatype, declare-datatypes, Int, false, forall},
	escapeinside     = {(*}{*)},
	backgroundcolor=\color{lbcolor},
	tabsize=4,    
	basicstyle=\scriptsize,
	upquote=true,
	aboveskip={1.5\baselineskip},
	columns=fixed,
	showstringspaces=false,
	extendedchars=false,
	prebreak = \raisebox{0ex}[0ex][0ex]{\ensuremath{\hookleftarrow}},
	frame=single,
	numbers=left,
	showtabs=false,
	showspaces=false,
	showstringspaces=false,
	identifierstyle=\ttfamily,
	keywordstyle={\ttfamily\color[rgb]{0,0,1}},
	commentstyle=\color{burgundy},
	stringstyle=\color[rgb]{0.627,0.126,0.941},
	numberstyle=\color[rgb]{0.205, 0.142, 0.73},
    float=tb,
    floatplacement=tb,
    abovecaptionskip=-20pt,
    belowskip=-15pt
}
\lstdefinelanguage{SMT-Prolog} {
	alsoletter       = {-, [=], :},
	morecomment      = [l]{;},
	morekeywords     = {define-fun, declare-const, declare-heap,
		Bool,
		bvadd, bvsub, bvneg, bvsgt, bvsle,
		select, store, assert, and, or, not, let,
		declare-datatype, declare-datatypes, Int,
	    :-, emptyHeap, allocate, write, read, valid, true, false, newAddress, newHeap, alloc},
	escapeinside     = {(*}{*)},
	backgroundcolor=\color{lbcolor},
	tabsize=4,    
	basicstyle=\scriptsize,
	upquote=true,
	aboveskip={1.5\baselineskip},
	columns=fixed,
	showstringspaces=false,
	extendedchars=false,
	prebreak = \raisebox{0ex}[0ex][0ex]{\ensuremath{\hookleftarrow}},
	frame=single,
	numbers=left,
	showtabs=false,
	showspaces=false,
	showstringspaces=false,
	identifierstyle=\ttfamily,
	keywordstyle={\ttfamily\color[rgb]{0,0,1}},
	commentstyle={\ttfamily\color{burgundy}},
	stringstyle=\color[rgb]{0.627,0.126,0.941},
	numberstyle=\color[rgb]{0.205, 0.142, 0.73},
} 

\newcommand{\mysubsubsection}[1]{\vspace{-2ex}\subsubsection{#1}}
\newcommand{\princess}{\textsc{Prin\-cess}}
\newcommand{\eldarica}{\textsc{Eld\-arica}}

\title{A Theory of Heap for Constrained Horn Clauses\\(Extended Technical Report)}
\titlerunning{A Theory of Heap for CHCs (Extended Technical Report)}
\author{Zafer Esen \orcidID{0000-0002-1522-6673} \and 
        Philipp R\"ummer \orcidID{0000-0002-2733-7098} }
\institute{Uppsala University, Sweden}

\begin{document}
	\newcommand{\cmt}[1]{#1}
	\newcommand{\fun}[1]{\textrm{$\mathsf{#1}$}}
	\newcommand{\rd}{\fun{read}}
	\newcommand{\wt}{\fun{write}}
	\newcommand{\ia}{\fun{valid}}
	\newcommand{\eh}{\fun{emptyHeap}}
	\newcommand{\ct}{\fun{heapSize}}
	\newcommand{\nh}{\fun{newHeap}}
	\newcommand{\na}{\fun{newAddr}}
	\newcommand{\ntha}{\fun{nthAddress}}
	\newcommand{\alloc}{\fun{allocate}}
	\newcommand{\defObj}{\mathit{defObj}}
	\newcommand{\allocRes}{AllocationResult}
	\newcommand{\nullAddr}{\fun{nullAddress}}
    \newcommand{\fr}{\fun{free}}
	\newcommand{\all}[1]{\forall #1.}
	\newcommand{\fls}{\mathit{false}}
	\newcommand{\tru}{\mathit{true}}
	\newcommand{\sel}{\mathit{select}}
	\newcommand{\str}{\mathit{store}}
	\newcommand{\addr}{\mathit{Address}}
	\newcommand{\hp}{\mathit{Heap}}
	\newcommand{\obj}{\mathit{Object}}
	\newcommand{\Int}{\mathit{Int}}
	\newcommand{\Bool}{\mathit{Bool}}
	\newcommand{\Nat}{\mathit{Nat}}
	\newcommand{\pair}[2]{\langle #1,#2 \rangle}
        \newcommand{\axiomref}[1]{[\ref{#1}]} 
\maketitle

\renewcommand{\thelstlisting}{\arabic{lstlisting}}

\begin{abstract}
  Constrained Horn Clauses (CHCs) are an intermediate program
  representation that can be generated by several verification tools,
  and that can be processed and solved by a number of Horn
  solvers. One of the main challenges when using CHCs in verification
  is the encoding of \emph{heap-allocated data-structures:} such
  data-structures are today either represented explicitly using the
  theory of arrays, or transformed away with the help of
  invariants or refinement types, defeating the purpose of CHCs as a
  representation that is language-independent as well as agnostic of
  the algorithm implemented by the Horn solver.  This paper presents
  an \emph{SMT-LIB theory of heap} tailored to CHCs, with the goal of
  enabling a standard interchange format for programs with heap
  data-structures. We introduce the syntax of the theory of heap,
  define its semantics in terms of axioms and using a reduction to
  SMT-LIB arrays and data-types, and discuss its properties and outline
  possible extensions and future work.
\end{abstract}

\section{Introduction} \label{sec:intro}
Constrained Horn Clauses (CHCs) are a convenient intermediate
verification language that can be generated by several verification
tools in many settings, ranging from verification of smart contracts
\cite{DBLP:conf/ndss/KalraGDS18} to verification of computer programs
in various languages
\cite{DBLP:conf/pldi/GrebenshchikovLPR12,seahorn,jayhorn-2017,rusthorn,DBLP:conf/pepm/SatoI019}. The
CHC interchange language provides a separation of concerns, allowing
the designers of verification systems to focus on high-level aspects like
the applied proof rules and verification methodology, while giving CHC
solver developers a clean framework that can be instantiated using
various model checking algorithms and specialised decision
procedures. Solver performance is evaluated in the annually held
CHC-COMP~\cite{DBLP:journals/corr/abs-2008-02939}.

CHCs are usually expressed using the SMT-LIB standard, which itself is
a common language and interface for SMT solvers
\cite{smt-2.6}. Abstractly, both SMT solvers and CHC solvers are tools
that determine if a first-order formula is satisfiable modulo
background theories such as arithmetic, bit-vectors, or arrays.

One of the main challenges when using CHCs, and in verification in
general, is the encoding of programs with mutable, heap-allocated
data-structures. Since there is no native theory of heap in SMT-LIB,
one approach to represent such data-structures is using the theory of
arrays (e.g.,
\cite{DBLP:conf/fmcad/KomuravelliBGM15,DBLP:journals/fuin/AngelisFPP17a}). This
is a natural encoding since heap can be seen as an array of memory
locations; however, as the encoding is byte-precise, in the
context of CHCs it tends to be low-level and often yields
clauses that are hard to solve.

An alternative approach is to transform away such data-structures with
the help of invariants or refinement types (e.g.,
\cite{DBLP:conf/pldi/RondonKJ08,DBLP:conf/sas/BjornerMR13,DBLP:conf/sas/MonniauxG16,jayhorn-2017}). In
contrast to approaches that use the theory of arrays, the resulting
CHCs tend to be over-approximate (i.e., can lead to false positives),
even with smart refinement strategies that aim at increasing
precision. This is because every operation that reads, writes, or
allocates a heap object is replaced with assertions and assumptions
about local object invariants, so that global program invariants might
not be expressible. In cases where local invariants are sufficient,
however, they can enable efficient and modular verification even of
challenging programs.


Both approaches leave little design choice with respect to handling of
heap to CHC solvers. Dealing with heap at encoding level implies
repeated effort when designing verifiers for different programming
languages, makes it hard to compare different approaches to encode
heap, and is time-consuming when a verifier wants to switch to another
encoding. The benefits of CHCs are partly negated, since the discussed
separation of concerns does not carry over to heap.

The vision of this paper is to extend CHCs to a standardised
interchange format for programs with heap data-structures. To this
end, we present a high-level theory of heap that does not restrict the
way in which CHC solvers approach heap, while covering the main
functionality of heap needed for program verification:
(i)~representation of the type system associated with heap data;
(ii)~reading and updating of data on the heap;
(iii)~handling of object allocation.

We use algebraic data types (ADTs), as already standardised by SMT-LIB
v2.6, as a flexible way to handle (i). The theory offers operations
akin to the theory of arrays to handle (ii) and (iii). The theory is
deliberately kept simple, so that it is easy to add support to SMT and CHC
solvers: a solver can, for instance, internally encode heap using
the existing theory of arrays (we provide one such encoding in
Section~\ref{sec:encoding-array}), or implement transformational
approaches like
\cite{DBLP:conf/sas/BjornerMR13,DBLP:conf/sas/MonniauxG16}. Since we
want to stay high-level, arithmetic operations on pointers are
excluded in our theory, as are low-level tricks like extracting
individual bytes from bigger pieces of data through pointer
manipulation. Being language-agnostic, the theory of heap allows for
common encodings across different applications, and is in the spirit
of both CHCs and SMT-LIB.

\begin{table}[b]
\begin{lstlisting}[label=lst-motivating-ex,caption=The motivating example in Java, abovecaptionskip=-20pt]
\end{lstlisting}\vspace{-20pt}
\begin{minipage}{0.46\textwidth}
\begin{lstlisting}[style=myJava,firstline=1,lastline=16]
abstract class IntList {
  protected int _sz;
  abstract int hd();
  abstract void setHd(int hd);
  abstract IntList tl();
  int sz() {return _sz;} }

class Nil extends IntList {
  Nil() {_sz = 0;}
  int hd () {err();}
  void setHd (int hd) {err();}
  IntList tl () {err();} }

class Cons extends IntList {
  int _hd;
  IntList _tl;
\end{lstlisting}
\end{minipage}\hfill
\begin{minipage}{0.46\textwidth}
\begin{lstlisting}[style=myJava,firstline=17,firstnumber=17]
  int hd() {return _hd;}
  void setHd (int hd) {_hd=hd;}
  IntList tl() { return _tl; }
  Cons(int hd, IntList tl) {
    _hd = hd;
    _tl = tl;
    _sz = 1 + tl.sz(); } }
class Motivation {
  void main() {
    IntList l = new Cons(42,
                  new Nil());
    l.setHd(l.hd()+1);
    assert(l.hd() == 43);
  }
}
\end{lstlisting}
\end{minipage}
\end{table}

\paragraph{Contributions of the paper} are
\begin{inparaenum}[(i)]
\item the definition of syntax and two possible formulations of
  semantics of the theory of heap (axiomatic and through an encoding into the theory of arrays);
\item a discussion on how programs can be
  encoded using the heap theory; and
\item properties of the theory.
\end{inparaenum}

\paragraph{Acknowledgements.}

This is the first full paper introducing the theory of heap.  An
earlier version of the theory was presented at the HCVS
Workshop~2020~\cite{heapTheoryHCVS} and the SMT Workshop~2020. An
invited paper at LOPSTR~2020 discusses preliminary work on decision
and interpolation procedures~\cite{10.1007/978-3-030-68446-4_9}.  We
are grateful for the discussion and feedback provided by the different
communities.

\section{Motivating Example}
\label{sec:example}

We start with a high-level explanation how heap is handled by our theory.
%
Listing~\ref{lst-motivating-ex} shows a simple Java program which constructs a singly-linked list, highlighting various heap interactions such as allocating objects on the heap (lines 26--27), as well as reading (lines 28--29) and modifying (line 28) heap data. 

In order to encode this program we use Constrained Horn Clauses (CHCs), which we assume knowledge of (see Section~\ref{subsec:CHCs} for a brief introduction). Although we present the theory in the context of CHCs,
there is nothing CHC-specific in the theory itself; as discussed earlier
support for the theory can easily be added to both SMT and CHC solvers since it is kept deliberately high level and simple. The encoding is given in Listing~\ref{lst-encoding-java} in SMT-LIB v2.6 format.  

\vspace{-10pt}
\subsubsection{Heap declaration} \label{subsec:heap-declaration}
To encode this program using the theory of heap, first a heap has to
be declared that covers the program types as shown at lines 1--12 of
Listing~\ref{lst-encoding-java}. Each heap comes with its own sorts
for the heap itself and for heap locations (or addresses). Lines 2
and 3 are the names of declared heap and address sorts. We next need
to define which data can be placed on the heap, which is done by
choosing the sort of heap objects; this sort can be any of the sorts
declared prior to or together with the heap declaration, excluding the
heap sort itself. Line 4 specifies the object sort to be the ADT
\verb!Object!, declared later.


Line~5 defines the object assumed to be stored at \emph{unallocated}
heap locations. Since functions in SMT-LIB are total, semantics has to
be defined also for reads from such unallocated addresses. The theory
of heap leaves the choice of object produced by such reads to the
user; the term specified at line 5 must have the object sort chosen at
line 4. We call this the default object (or $\defObj$), which in this
case is created using the object constructor \texttt{O\_Empty}.

The rest of the heap declaration at lines 6--12 corresponds to an
SMT-LIB data-type declaration. In line 6, in addition to \verb!Object!
we declare data-types \verb!IntList!, \verb!Cons!, and \verb!Nil!,
encoding the classes of the program. The constructors at lines 7--9
specify the fields of each class, and in addition give \texttt{Cons}
and \texttt{Nil} each a field containing the parent \texttt{IntList}
object. In lines 10--12, the constructors of the \verb!Object! sort are
declared, which correspond to the classes \verb!Cons! and \verb!Nil!,
as well as the default object \verb!O_Empty!. The class \verb!IntList!
is abstract and does not occur directly on the heap, so that no
constructor for this type is provided.

Since each heap theory has its own address sort, cases are immediately
prevented in which multiple heaps share the same address sort, or in
which some other interpreted sort (say, \verb!Int!) is used to store
addresses. This rules out accidental cases of pointer arithmetic, and
leaves full flexibility to solvers on how to internally represent
addresses (e.g., see \cite{jayhorn-2017}). This choice also makes it
necessary to include the ADT declarations within \verb!declare-heap!,
since ADTs representing objects often have to refer to the address
sort.

Within one heap, all pointers are represented using a single $\addr$
sort, and no distinction is made between pointers to objects from
different constructors. This is close in semantics to languages
like C, where casts between arbitrary pointer types are possible, and
it has to be verified for each heap access that indeed an object of
the right type is accessed. In languages like Java, the stronger type
system will provide information about the objects a variable can refer
to, but exceptions can be raised when performing casts. The theory of
heap is flexible enough to cover those different settings.



Apart from the sorts mentioned, the heap declaration implicitly
declares an ADT \texttt{ARHeap} (also called $\allocRes\hp$ later in the paper) 
that holds pairs $\langle \hp,\addr \rangle$ returned as a result of allocations.

\vspace{-10pt}
\subsubsection{Program encoding}
\newcommand{\betweenColor}[3]{\ifnum
         \value{lstnumber}>#1
            \ifnum\value{lstnumber}<#2
                \color{#3}
            \fi
        \fi}

\begin{lstlisting}[language=SMT-LIB, caption=SMT-LIB encoding of the motivating example from Listing~\ref{lst-motivating-ex}. The symbols of some sorts and operations of the theory are abbreviated and the list of quantified variables are skipped in some cases for brevity.,label=lst-encoding-java,linebackgroundcolor={%
  \betweenColor{0}{19}{lbcolor}
  \betweenColor{18}{20}{lbcolor!50}
  \betweenColor{19}{23}{lbcolor}
  \betweenColor{22}{27}{lbcolor!50}
  \betweenColor{26}{29}{lbcolor}
  \betweenColor{28}{32}{lbcolor!50}
  \betweenColor{31}{34}{lbcolor}
  \betweenColor{33}{37}{lbcolor!50}
  \betweenColor{36}{39}{lbcolor}
}
]
(declare-heap
 Heap                                      ; declared Heap sort
 Addr                                      ; declared Address sort
 Object                                    ; chosen Object sort
 O_Empty                                   ; the default Object
 ((IntList 0) (Cons 0) (Nil 0) (Object 0)) ; ADTs
 (((IntList   (sz         Int)))           ; Class constructors
  ((Cons      (parentCons IntList) (hd Int) (tl Addr)))
  ((Nil       (parentNil  IntList)))
  ((O_Cons    (getCons    Cons))           ; Object sort constructors
   (O_Nil     (getNil     Nil))
   (O_Empty   ))))
                                           ; invariant declarations
(declare-fun I1 (Heap)      Bool)          ; <h>
(declare-fun I2 (Heap Addr) Bool)          ; <h,p>
(declare-fun I3 (Heap Addr) Bool)          ; <h,l>
(declare-fun I4 (Heap Addr) Bool)          ; <h,l>

(assert (I1 emptyHeap))
(assert (forall ((h Heap) (h1 Heap) (p1 Addr))
  (=> (and (I1 h) (= (ARHeap h1 p1) (alloc h (O_Nil (Nil (IntList 0))))))
      (I2 h1 p1))))
(assert (forall (...)
  (=> (and (I2 h p)
           (= (ARHeap h1 p1) (alloc h (O_Cons (Cons (IntList 1) 42 p)))))
      (I3 h1 p1))))
(assert (forall (...)
  (=> (and (I3 h l) (not (valid h l))) false)))
(assert (forall (... (pn IntList) (head Int) (tail Addr))
  (=> (and (I3 h l) (= h1 (write h l (O_Cons (Cons pn (+ 1 head) tail))))
           (= (O_Cons (Cons pn head tail)) (read h l))) (I4 h1 l))))
(assert (forall (...)
  (=> (and (I3 h l) (= (O_Nil (Nil pn)) (read h l))) false)))
(assert (forall (...)
  (=> (and (I4 h l) (= (O_Cons (Cons pn head tail)) (read h l))
           (not (= head 43))) false)))
(assert (forall (...)
  (=> (and (I4 h l) (not (is-O_Cons (read h l)))) false)))
\end{lstlisting}

Invariants representing program states are declared at lines 14--17. The first set of arguments in the parentheses list the sorts of the variables we want to keep track of at that point. E.g., for line 17, we want to have a global view of the heap, as well as all variables on the stack at that point. The only variable on the stack at this point is a temporary variable \verb!p! that corresponds to the newly allocated \verb!Nil! object's address (line 27 in Listing~\ref{lst-motivating-ex}). 

Line 19 is the program entry point, where the heap is initially
empty. The function $\eh$ returns an empty heap (i.e., unallocated at
all locations) of the declared $\hp$ sort specified at line 2.  Lines
20--26 allocate, respectively, a \texttt{Nil} object and a
\texttt{Cons} object on the heap. Allocation is done using the $\alloc$ 
function of the theory, which takes as arguments the old heap and the
new object to be put on the heap, and returns an \verb!ARHeap!  pair
with the new heap and the allocated address. Constructor calls are
inlined and slightly simplified in the encoding. For example, line 25
shows the simplified encoding of the Java constructor for
\texttt{Cons} at lines 20--23 of Listing~\ref{lst-motivating-ex}. The
updating of the \texttt{\_sz} field is simplified by directly
assigning a value to it, which would actually require another clause
with a read due to the statement at line 23 of
Listing~\ref{lst-motivating-ex}.

Lines 27--33 correspond to the statement at line 28 from
Listing~\ref{lst-motivating-ex}, which calls the methods \texttt{hd}
and \texttt{setHd} corresponding to a read-modify-write operation on
the list. We again inline these methods in the encoding; however,
since both \texttt{Nil} and \texttt{Cons} define these methods, we add
a clause for each (lines 29--31 encode \texttt{Cons.hd()} and
\texttt{Cons.setHd()}, while lines 31--33 encode
\texttt{Nil.hd()}). For brevity we do not show the clause for the inlined call to \texttt{Nil.setHd()}, which is similar to the
encoding at lines 32--33. The assertion at lines 27--28 checks the validity of accesses in order to ensure memory safety.

Lines 27--33 illustrate the use of $\rd$ and $\wt$ functions. $\rd$ 
reads from the provided heap at the given location, and
$\wt$ writes the provided object to the heap at the
specified location. The dynamic dispatch needed when calling \verb!hd!
is implemented through pattern matching using the \texttt{O\_Cons} and
\texttt{O\_Nil} constructors: in lines 29--31 the method call is
successful, and the heap object is subsequently updated, while the
clause at lines 32--33 models the error when executing
\verb!Nil.hd!. The same property can be expressed using
the tester \texttt{is-O\_Cons} in lines~37--38.
%
%
Lastly, lines 34--36 encode the assertion at line 29 from
Listing~\ref{lst-encoding-java}.

\section{Preliminaries}
\subsubsection{Definition of a Theory}
A \emph{signature} (or vocabulary) $\Sigma$ of a many-sorted logic is
defined as the triple containing a set $S$ of sorts, a set $\Sigma_f$
of function symbols, and a set $\Sigma_p$ of relation symbols. The
arguments of functions and relations, and the values of functions are
specified using sorts from $S$. A $\Sigma$-\emph{formula} uses only
non-logical symbols from $\Sigma$, in addition to logical symbols.
A $\Sigma$-\emph{sentence} is a $\Sigma$-formula that contains no free
variables.

A $\Sigma$-\emph{theory} $T$ is defined as a set of $\Sigma$-sentences
closed under entailment. A $\Sigma$-formula $\phi$ is said to be
$T$-satisfiable if a structure exists which satisfies both the
sentences of $T$ and $\phi$. We call this structure a $T$-\emph{model} of
$\phi$.
\mysubsubsection{The Theory of Arrays} \label{subsec:arrays}
The idea of a (non-extensional) first order theory of arrays was first introduced by McCarthy \cite{DBLP:conf/ifip/McCarthy62}. It has the two functions $\sel$ and $\str$, whose semantics are given through the following read-over-write axioms:
\newtagform{brackets}{[}{]}
\usetagform{brackets} 
\begin{align}
\forall a, i, j, e. (i = j &\rightarrow \sel(\str(a, i, e), j) = e) \label{eq:array-row-1}\tag{array-row1}\\ 
\forall a, i, j, e. (i \neq j &\rightarrow \sel(\str(a, i, e), j) = \sel(a,j)) \label{eq:array-row-2}\tag{array-row2}
\end{align}
where $a$ is an array, $i$ and $j$ are indices, and $e$ is an element stored in the array.

Extensionality is introduced by an additional axiom, which allows reasoning about equality between two arrays:
\begin{align}
\forall a_1, a_2, i. (\sel(a_1,i) = \sel(a_2,i) \rightarrow a_1 = a_2) \label{eq:array-ex}\tag{array-ex}
\end{align}
The theory of arrays is one of the background theories defined by SMT-LIB, and as such, many solvers have specialised decision procedures for the decidable fragments of this theory. 
\usetagform{default}
\mysubsubsection{Algebraic Data-Types (ADTs)}
Algebraic data-types (ADTs or data-types) provide a flexible way to represent types in many programming languages, and many SMT solvers provide native decision procedures to solve them efficiently \cite{cvc4,princess08,z3}. They are supported in the SMT-LIB standard since version 2.6 through the \texttt{declare-datatype} and \texttt{declare-datatypes} commands. Non-recursive ADTs can be used to represent programming types such as enumerations, records and unions, while recursive ADTs can represent types such as arrays, lists and strings. 

\mysubsubsection{Constrained Horn Clauses (CHCs)} \label{subsec:CHCs}
A CHC is a sentence
\[\forall{x_1,x_2,...}.~\big(C \wedge B_1 \wedge ... \wedge B_n \rightarrow H\big)\]
where $H$ is either an application of a $k$-ary predicate $p(t_1,...,t_k)$ to first-order terms or $\mathit{false}$, $B_i$ (for $i=1 \ldots n$) is an application of an $m$-ary predicate $p_i(t_1,...,t_m)$ to first-order terms, and $C$ is a constraint over some background theories (in this case including the proposed theory of heap). The universal quantification of first-order variables in a clause is usually not explicitly specified. 

CHCs provide a natural way to encode programs: invariants represent program states, state transitions and assertions can be encoded through constraints and contradictions. A set of CHCs is solvable if no contradiction can be derived. We refer to other sources such as \cite{DBLP:conf/birthday/BjornerGMR15,DBLP:conf/pldi/GrebenshchikovLPR12} for a more comprehensive explanation.

\section{Vocabulary and Syntax of the Theory of Heap}
\label{sec:syntax}
\subsection{SMT-LIB-style Declaration of Heaps}
A theory of heap is declared as follows:
\[\boxed{(\texttt{declare-heap}~
 c_h~c_a~c_o~\tau_o~
 ((\delta_1 k_1)~...~(\delta_n k_n))~
 (d_1 ... d_n)
)}\]
where $c_h,~c_a,~c_ o$ are symbols corresponding to the names of declared heap,
declared address and chosen object respectively. $\tau_o$ is a term of the
chosen object which is returned on invalid accesses (i.e. the default object).
The object sort can be chosen as any sort except $c_h$. The rest of the
declaration resembles the \verb!declare-datatypes! declaration from the SMT-LIB
standard v2.6 \cite{smt-2.6}, with the exception that polymorphism is
(currently) not supported in constructor declarations, and that there should
be $n$ (where $n \geq 0$) instead of $n+1$ ADT sort declarations (i.e., 
the object sort can also be declared before the heap declaration and specified
using $c_o$, if it does not use the address sort ($c_a$) in its declaration).

\newcommand{\bnfg}[1]{\textcolor{color2}{\mathit{\langle #1 \rangle }}}
The concrete syntax for the heap declaration is given below, which extends 
$\bnfg{command}$ in the concrete syntax of SMT-LIB v2.6.

\begin{tabular}{lcl}
$\bnfg{command}$ &    ::=    &...\\
                 &  $\vert$  &$ (~\texttt{declare-heap}~
 \bnfg{symbol}~\bnfg{symbol}~\bnfg{sort}~\bnfg{term}$\\
  &&$(~\bnfg{sort\_dec}\textcolor{color2}{^{n}}~)~(~\bnfg{heap\_datatype\_dec}\textcolor{color2}{^{n}}~)~)$\\
$\bnfg{heap\_datatype\_dec}$ & ::= & $\bnfg{constructor\_dec}\textcolor{color2}{^+}$\\
\end{tabular}\\

The first two symbols and the following sort in the declaration correspond respectively to $c_h$, $c_a$ and $c_o$ from the abstract syntax. $\bnfg{term}$ is the
default object.

\subsection{Sorts}
Each heap declaration introduces several sorts. The names of these sorts are defined by the variables in the \verb!declare-heap! command, which we assume in this paper to be $\hp$ for $c_h$ and $\addr$ for $c_a$: 
\begin{itemize}
\item a sort $\hp$ of heaps,
\item a sort $\addr$ of
heap addresses,
\item zero or more ADT sorts used to represent heap data,
\item an additional ADT sort that holds the pair $\langle \hp, \addr \rangle$ which is the result of calling $\alloc$. In order to make this ADT sort distinguishable, it is suffixed with associated heap sort $\hp$ (e.g. $\allocRes\hp$).
\end{itemize} 

\subsection{Operations}\label{subsec:functions}
Below we describe each function of the theory; the semantics are given more formally through the axioms in Section~\ref{subsec:axioms}. They are not listed below, but we also get access to all ADT operations as a side effect of heap declarations. Some operations contain the symbols \verb!Heap! and \verb!Address! in their signatures. This is done with the assumption that the declared heap and address sorts are named $\hp$ and $\addr$ respectively. E.g. \verb!nullAddress! would be \verb!nullA! if the declared address sort was named $A$, and it would return the sort $A$. Including the sort name in some function and sort names makes it possible to determine their associated heap declarations without using the SMT-LIB command ``\verb!as!''. This is not required in sorts and functions where the associated heap sort is clear, such as in $\rd$ (its first argument is of heap sort).
\begin{empheq}[box=\fbox]{equation*}
    \nullAddr : () \to \addr
\end{empheq}
Function $\nullAddr$ returns an $\addr$ which is always unallocated/invalid.
\begin{empheq}[box=\fbox]{equation*}
    \eh : () \to \hp
\end{empheq}
$\eh$ returns the $\hp$ that is unallocated everywhere. 
\begin{empheq}[box=\fbox]{equation*}
    \alloc : \hp \times \obj \to \hp \times \addr~~ (\allocRes\hp)
\end{empheq}
Function
\alloc~takes a $\hp$ and an $\obj$, and returns $\allocRes\hp$. $\allocRes\hp$~is a data-type representing the pair $\langle \hp, \addr \rangle$. The returned $\hp$ at $\addr$  contains the passed $\obj$, with all other locations unchanged. The pair ADT is required as the return sort since it is not possible in SMT-LIB to return the two sorts separately. In Section~\ref{sec:extensions} we discuss other alternatives such as using multiple allocation functions.

\begin{empheq}[box=\fbox]{equation*}
    \ia : \hp \times \addr \to \mathit{Bool}
\end{empheq}
The predicate $\ia$ checks if accesses to the given $\hp$ at the given $\addr$ are valid. We say that an access is valid if and only if that location was allocated beforehand by using the function $\alloc$. 
\begin{empheq}[box=\fbox]{equation*}
    \rd : \hp \times \addr \to \obj
  \end{empheq}
  \vspace*{-4ex}
\begin{empheq}[box=\fbox]{equation*}
    \wt : \hp \times \addr \times \obj \to \hp
\end{empheq}
Functions \rd~and~\wt~are similar to the array \emph{select} and \emph{store} operations described in Section~\ref{subsec:arrays}; however, unlike an array, a heap also carries information about allocatedness. This means the \rd~and \wt~functions only behave as their array counterparts if the considered address is allocated. If the read address is unallocated, a default $\obj$ is returned to make the function total (as explained in Section~\ref{subsec:heap-declaration} / Heap Declaration).

The function \wt~normally returns a new $\hp$ if the access is valid. If not, then the original $\hp$ is returned without any changes. Validity of a \wt~can be checked via memory-safety assertions as shown in lines 27--28 of Listing~\ref{lst-encoding-java}.
\begin{empheq}[box=\fbox]{equation*}
    \ntha_i
\end{empheq}
We propose a further short-hand notation $\ntha_i$, which is useful when presenting satisfying assignments. It is used to concisely represent $\addr$ values which would be returned after $i$ \alloc~calls, which is only possible with the deterministic allocation axiom \axiomref{eq:alloc2} given in Section~\ref{subsec:axioms}.


\section{Encoding of Different Programming Languages}
\label{sec:encoding}

\subsubsection{Java and Java-like Languages}

We have outlined in Section~\ref{sec:example} how a Java class
hierarchy can be encoded using the theory of heap, and how the
different Java instructions can then be translated to CHCs.  Every
class is mapped to one ADT, representing inheritance by adding a
\emph{parent} field to the sub-classes of a class, and defining an
\verb!Object! ADT as the union of the types that can occur on the
heap. Java interfaces do not have to be considered explicitly, since
in Java they are abstract and do not store data. Arrays and strings
can in principle be handled using recursive ADTs, although it is
probably more efficient to integrate the theory of arrays for
this purpose (Section~\ref{sec:extensions}).
Java also supports parametric polymorphism (generics), but implements it
using type erasure, which means that type parameters do not explicitly
occur on the heap and do not have to be stored. In languages with
native polymorphism, for instance C\#, types can be encoded using
dedicated ADTs as part of a heap declaration, and type parameters of
classes and methods can be represented using explicit fields/arguments.

\mysubsubsection{Programs in C and C++}

Our theory implements a relatively abstract view of the heap, and does
not provide a byte-level heap model, which implies that not all C
features can be handled directly. We believe that the theory
represents a good trade-off, however, for analysing functional aspects
of C programs that avoid undefined behaviour.  Structs, enums, and
unions in C can all be mapped to ADTs in a similar way as Java
classes. C can in addition store native types like \verb!int! on the
heap, which can be encoded easily through further ADTs. Unsafe pointer
conversions can be supported by verifying, using appropriate CHCs,
that read/write accesses to objects only happen through the correct
type; CHCs can also define certain byte-level conversions of
objects. This way it is possible, among others, to give correct
semantics to patterns like byte-level heap allocation using
\texttt{malloc} or \texttt{calloc}.

Several other C features cannot be supported within the heap theory.
The theory strictly rules out pointer arithmetic between objects at
different addresses; it would be possible, however, to encode pointer
arithmetic within an object already at the encoding level. Stack
pointers are outside of the scope of the heap theory, but can also to
some degree be handled during the CHC encoding.  A further operation
allowed by C, but not considered in the theory, is the deallocation of
heap locations; this could be supported with the addition of a
\texttt{free} function to the theory. Extensions to the theory
are discussed in Section~\ref{sec:extensions}.

Heap in C++ can be modelled essentially by combining the techniques
discussed for Java and C. Multiple inheritance of classes, which is
possible in C++, can be encoded by adding multiple \emph{parent}
fields in the sub-classes. C++ templates, realising compile-time
polymorphism, can be handled by adding separate ADTs for each template
instance.

\section{Semantics of the Theory}\label{sec:semantics}

\subsection{Axiomatic Semantics}
\label{subsec:axioms}
We first propose and discuss a set of axioms defining the
semantics of the heap theory.
All variables occurring in the axioms are universally quantified with sorts $h: \hp$, $p: \addr$, $o: \obj$ and $\mathit{ar}: \allocRes\hp$. Variables can also appear subscripted. $\allocRes\hp$ is the pair $\langle \hp, \addr \rangle$, we use $\mathit{ar.\_1}$ and $\mathit{ar.\_2}$ to select the $\hp$ and $\addr$ fields of $\mathit{ar}$, respectively.
\usetagform{brackets} 
\mysubsubsection{Array-like axioms}
	
\begin{empheq}[box=\fbox]{equation}
    \ia(h, p) \rightarrow \rd(\wt(h,p,o),p) = o \tag{row1}\label{eq:row1}
\end{empheq}
\axiomref{eq:row1} defines the semantics of reading from an allocated $\addr~p$, to which the last $\wt$ was the $\obj~o$. This is similar to the array read-over-write axiom \axiomref{eq:array-row-1}, but is only applied when the accessed location is valid.
%
\begin{empheq}[box=\fbox]{equation}
    p_1 \neq p_2 \rightarrow \rd(\wt(h, p_1, o), p_2) = \rd(h, p_2) \tag{row2}\label{eq:row2}
\end{empheq}
\axiomref{eq:row2} says that reading $\addr~p_2$ from a $\hp~h$ written at $\addr~p_1$ is the same as directly reading $\addr~p_2$ from $\hp~h$.  
Checking for validity here is not required due to the axiom \axiomref{eq:ivwt}, since invalid writes return the same heap.
%
\begin{empheq}[box=\fbox]{equation}
	(\forall{p:\addr.}(\ia(h_1,p) \leftrightarrow \ia(h_2,p))\land \rd(h_1,p) = \rd(h_2,p)) \rightarrow h_1 = h_2	\tag{ext}\label{eq:ext}
\end{empheq}
The extensionality axiom \axiomref{eq:ext} states that, given any $\addr~p$, if two $\hp$s have the same allocation state at $p$, and reads from $p$ return the same $\obj$ in both, then the two $\hp$s must be the same. This axiom differs from the extensionality axiom of the theory of arrays \axiomref{eq:array-ex} only with the validity checks.

\mysubsubsection{Axioms about allocation}
\begin{empheq}[box=\fbox]{equation}
	\alloc(h, o) = ar \rightarrow \rd(ar.\_1, ar.\_2) = o 
	\tag{roa1}\label{eq:roa1}
\end{empheq}
The axiom \axiomref{eq:roa1} states that reading from a $\hp$, using the $\addr$ returned from an allocation using $\hp~h$ and $\obj~o$, returns $o$.
%
\begin{empheq}[box=\fbox]{equation}
	\alloc(h, o) = ar~\land~p \neq ar.\_2 \rightarrow \rd(ar.\_1, p) = \rd(h, p)
	\tag{roa2}\label{eq:roa2}
\end{empheq}
The axiom \axiomref{eq:roa2} states that reading from a $\hp$ using an $\addr~p$ that is different than the $\addr$ returned from the allocation, which was done using $\hp~h$ and $\obj~o$, is the same as directly reading $p$ from $h$.
%
\begin{empheq}[box=\fbox]{equation}
	\makecell{\alloc(h,o) = ar \rightarrow
	\lnot \ia(h,ar.\_2) \land \ia(ar.\_1,ar.\_2)~\land\\ 
    (\forall{p:\addr.}(ar.\_2 \neq p \rightarrow  (\ia(h,p) \leftrightarrow \ia(ar.\_1,p))))}
	\tag{alloc1}\label{eq:alloc1}
\end{empheq}
\axiomref{eq:alloc1} states that an allocation takes a $\hp~h$ and an $\obj~o$, and returns a $\langle\hp,\addr\rangle$ pair which is allocated (i.e. \ia).
The returned $\addr$ must have been unallocated at $h$. The last conjunct in the axiom states that the validity of both $\hp$s differ only at the $\addr$ which was just allocated. 

\begin{empheq}[box=\fbox]{equation}
	\makecell{(\forall{p:\addr.}(\ia(h_1,p)\leftrightarrow\ia(h_2,p)))\rightarrow\\ \alloc(h_1,o_1).\_2 = \alloc(h_2,o_2).\_2}
	\tag{alloc2}\label{eq:alloc2}
\end{empheq}
The axiom \axiomref{eq:alloc2} ensures that the allocations are deterministic. If two $\hp$s are valid at the same $\addr$es (i.e., due to the same number of $\alloc$~calls), then allocating a new $\obj$ on either will return the same $\addr$.

\mysubsubsection{Axioms about validity}

\begin{empheq}[box=\fbox]{equation}
	\lnot \ia(h, p) \rightarrow \wt(h, p, o) = h
	\tag{ivwt}\label{eq:ivwt}
\end{empheq}
The axiom \axiomref{eq:ivwt} states that a write to an invalid $\addr$ of $\hp~h$ returns $h$, in other words, the heap is unchanged by invalid writes, which eliminates the need for a validity check on the left-hand side of the implication in \axiomref{eq:row2}.

\begin{empheq}[box=\fbox]{equation}
	\lnot \ia(h, p) \rightarrow \rd(h, p) =~\defObj
	\tag{ivrd}\label{eq:ivrd}
\end{empheq}
The axiom \axiomref{eq:ivrd} states that a read from an invalid $\addr$ of a $\hp$ returns the default $\obj$ i.e., $\defObj$ (as explained in Section~\ref{subsec:heap-declaration} / Heap Declaration).

\begin{empheq}[box=\fbox]{equation}
	\lnot \ia(\eh, p)
	\tag{vld1}\label{eq:vld1}
\end{empheq}
The axiom \axiomref{eq:vld1} states that $\eh$ is unallocated at every $\addr$. 

\begin{empheq}[box=\fbox]{equation}
	\lnot \ia(h, \nullAddr)
	\tag{vld2}\label{eq:vld2}
\end{empheq}
The axiom \axiomref{eq:vld2} states that $\nullAddr$ is unallocated in every $\hp$.

\subsubsection{No-junk (or constructability) axiom}
\begin{empheq}[box=\fbox]{equation}
	\makecell{\exists{f: \Nat \to \hp, g: \Nat \to \addr}.
			\\f(0) = \eh \land g(0) = \nullAddr~\land\\
\forall{i: \Nat}.~\langle f(i+1),g(i+1) \rangle = \alloc(f(i), \defObj)~\land\\
			~~\forall{p: \addr}.~\exists{i: \Nat.~g(i) = p}}
	\tag{cons}\label{eq:cons}
\end{empheq}
The axiom \axiomref{eq:cons} makes the $\hp$ constructable by enumerating every $\hp$ and $\addr$. It is required in order to ensure that there are no heap terms in the models which cannot be generated. $\defObj$ is used in this axiom as a generic object since the allocated object in this case is not of importance. 

\usetagform{default}

\subsection{Constructing a Model of the Axioms}\label{sec:encoding-array}

We now discuss how a model of the axioms can be defined in terms of
the theory of arrays. Such a reduction to arrays has multiple use
cases:
\begin{inparaenum}[(i)]
\item it witnesses consistency of the axioms;
\item in SMT solvers (but probably not in CHC solvers, as shown in our
  experiments below) it gives rise to a practical decision procedure;
  and
\item it enables us to carry over complexity results for the theory of
  arrays.
\end{inparaenum}

\begin{lstlisting}[language=SMT-LIB, label=lst-array-encoding, caption=Provisional encoding of the theory of heap using the theory of arrays]
(define-sort Addr  () Int)
(declare-datatypes <datatypes-from-heap-declaration>)
(declare-datatypes ((ARHeap 0) (Heap 0))
  (((ARHeap   (_1 Heap) (_2 Addr)))
   ((HeapCtor (heapSize Int) (contents (Array Addr Object))))))
(define-fun nullAddr  () Addr 0)
(define-fun defObj    () Object O_Empty)
(define-fun valid ((h Heap) (p Addr)) Bool
  (and (>= (heapSize h) p) (> p 0)))
(define-fun emptyHeap () Heap
  (HeapCtor 0 ((as const (Array Addr HeapObject)) defObj)))
(define-fun read      ((h Heap) (p Addr)) Object
  (ite (valid h p) (select (contents h) p) defObj))
(define-fun write     ((h Heap) (p Addr) (o Object)) Heap
  (ite (valid h p) (HeapCtor (heapSize h) (store (contents h) p o)) h))
(define-fun allocate  ((h Heap) (o Object)) ARHeap (ARHeap
  (HeapCtor (+ 1 (heapSize h)) (store (contents h) (+ 1 (heapSize h)) o))
  (+ 1 (heapSize h))))
(define-fun heap-eq   ((h1 Heap) (h2 Heap)) Bool (forall ((p Addr))
  (and (= (valid h1 p) (valid h2 p)) (= (read h1 p) (read h2 p)))))

\end{lstlisting}

The first attempt to define such a model is shown in
Listing~\ref{lst-array-encoding}.
The address sort $\mathit{Addr}$ is represented using integers, and ADT declarations that were previously part of a heap declaration are turned into a datatype declaration. Each $\hp$ term is associated with an array and an integer counter keeping track of the number of allocations (lines 6--7).

Each operation of the theory is then defined according to the axioms of the theory. $\ia$ becomes a simple check on $\ct$ and the integer value of the $\addr$ (lines 10--11). Line 12 declares an uninitialised array which is used to construct the $\eh$ on the next line. $\rd$ and $\wt$ operations become simple wrappers for array accesses, where the partial mapping is achieved using the \texttt{ite} (if-then-else) operator. $\alloc$ semantics are achieved by incrementing the $\ct$ after each allocation, and storing the allocated object at this location.

The encoding in Listing~\ref{lst-array-encoding} approximates heaps,
but still violates several of the heap axioms. Firstly, it does not
establish extensionality (axiom~[ext]), since array extensionality
does not exactly correspond to heap extensionality; the latter only
considers allocated addresses. This can be addressed by defining a
\texttt{heap-eq} predicate replacing negative occurrences
of the built-in equality~\verb!=! on heaps.

Secondly, the use of the sort \verb!Int! for addresses and heap size
is not consistent with the semantics stipulated by the
axioms. Negative addresses would describe memory locations that are
not reachable through $\alloc$, violating [cons], and
existence of heaps with negative heap size violate the axioms [roa1]
and [alloc1]. Since addresses can also be stored in heap objects,
fixing these issues by introducing additional well-formedness
constraints on the SMT-LIB level is cumbersome. The problems go away
when switching from \verb!Int! to the natural numbers \verb!Nat!,
which is not possible in SMT-LIB but easily doable solver-internally.

\section{Properties of the Theory of Heap}

The reduction shown in Section~\ref{sec:encoding-array} indicates that
basic properties of the theory of arrays carry over to the heap
theory, and in particular that satisfiability of quantifier-free heap
formulas is NP-complete (provided that the theory chosen to represent
heap objects is by itself in NP). Like for arrays, NP-completeness can
be observed already for conjunctions of heap literals.
 Proofs are in the appendix.
\begin{lemma}\label{lem:NP}
  Consider an instance of the heap theory with uninterpreted object
  sort~$O$. It is an NP-complete problem to check satisfiability of
  formulas~$\phi_1 \wedge \cdots \wedge \phi_n$, in which each
  $\phi_i$ is
  \begin{inparaenum}[(i)]
  \item an equation between terms involving variables and the
    functions $\rd$, $\wt$, $\alloc$, $\nullAddr$, $\eh$; or
  \item an atom $\ia(h, p)$; or
  \item the negation of an atom as in (i) or (ii).
  \end{inparaenum}
\end{lemma}

\begin{lemma}\label{lem:interpolation}
  The theory of heap does not admit quantifier-free Craig
  interpolation: there are unsatisfiable quantifier-free conjunctions
  $A \land B$ that do not have quantifier-free interpolants.
\end{lemma}

\section{Alternative Definitions and Extensions}
\label{sec:extensions}

This section explains the rationale behind some of the design choices
in the theory of heap, as well as some natural
extensions. It is intended as a starting point for further
discussions and a standardisation within SMT-LIB.

\mysubsubsection{$\allocRes\hp$}

Allocation on the heap needs to produce both a new heap and a fresh
address. In our theory, the pair of new heap and new address is
handled using the ADT $\allocRes\hp$, which enables us to stick to
just a single allocation function~$\alloc$. Alternatively, $\alloc$
could be represented using a pair of functions, as in
$\alloc(h,o) = \langle \fun{allocHeap}(h,o), \fun{allocAdress}(h,o)
\rangle$; this would be preferable from a solver implementation point
of view, but not necessarily for users. Altogether this point is more
of aesthetic concern.

\mysubsubsection{Deterministic allocation}

In the current semantics of the heap theory, object allocation is
deterministic: since $\alloc$ is a function, it will always produce
the same fresh address when applied to the same arguments.  Moreover,
\axiomref{eq:alloc2} implies that the new address is determined
entirely by the set of already allocated addresses on the heap.
Determinism is required for constructability of heaps, and for
presentation of counterexamples. It also simplifies the computation of
program invariants, since it implies the existence of a linear order
of the heap addresses, as witnessed by the array semantics in
Section~\ref{sec:encoding-array}:  an invariant can distinguish fresh
and used addresses using a simple inequality. Determinism will in many
practical cases not be observable in programs: the syntax of the heap
theory prevents arithmetic on addresses, and normal program semantics
does not allow $\alloc$ to be called repeatedly on the same heap in any
case.

In cases where it is needed, there is an elegant way to reintroduce
non-determinism: the $\alloc$ function can be given a third \emph{entropy}
argument, as in $\alloc(h, o, e)$, and the axiom \axiomref{eq:cons} be
relativized to only hold for fixed values of $e$. The axiom
\axiomref{eq:alloc2} could be dropped. The translation of programs to
CHCs can then choose a non-deterministic value for $e$ when encoding
an allocation operation like \verb!new!.  A side effect of this change
would be that decision procedures and correct encoding of
heap using arrays become more complex, and for instance have to store
the allocation status of each address using a bit-array.

\mysubsubsection{Deallocation}

A natural extension of the theory is the addition of a function for
deallocating objects, which would obviously be helpful to capture
languages without garbage collector, like C/C++; for such languages
deallocation otherwise has to be encoded using an explicit flag added
to objects. The effect on the theory semantics would be similar as for
non-deterministic allocation: decision procedures would need to
maintain an additional bit-array to remember the allocation status of
addresses.

\mysubsubsection{Integration of arrays}

A second relevant extension is the integration of the theory of heap
with McCarthy-style arrays. As defined here, the heap theory readily
allows arrays with primitive index and value-type to be stored as
objects on the heap. However, it is not possible to store \emph{arrays
  containing addresses,} in the same way in which the command
\begin{verbatim}
  (declare-datatype A ((a (x (Array Int A)))))
\end{verbatim}
is not a well-formed declaration in SMT-LIB. Storing addresses in
arrays on the heap probably does not pose any challenge for
implementing decision procedures, but requires a suitable
generalisation of the definitions in Section~\ref{sec:syntax}.

\mysubsubsection{Polymorphic Heap Objects}

The \verb!declare-heap! syntax is already prepared for including ADT
constructors with sort parameters (as in \verb!declare-datatypes!),
and in some cases parametric polymorphism would help to represent
class hierarchies of programs more succinctly. A full extension to
polymorphic heap objects requires further research, however, and the
overall value is not clear. Our experience is that the type systems in
intermediate verification languages, even when they provide
polymorphism~\cite{boogieTypeEncoding2010}, are often too weak to
directly capture the type systems of real-world programming languages
(with idiosyncratic sub-typing rules, native types
vs.\ boxed types, etc.), so that still an encoding is
necessary. Such an encoding can be done using ADTs in our theory
already now.

\section{Related Work}

\paragraph{Separation Logic} extends the assertions of Hoare's logic \cite{DBLP:journals/cacm/Hoare69} to succinctly express properties of heap and  shared mutable data-structures \cite{DBLP:conf/lics/Reynolds02}. Research has been done on specialised decision procedures for separation logic in SMT \cite{DBLP:conf/atva/ReynoldsIS016,DBLP:conf/aplas/PerezR13}, and there is a proposal for encoding separation logic in SMT-LIB 2.5 \cite{Iosif2018EncodingSL}.

The theory of heap and separation logic both provide mechanisms for reasoning about the heap; however, their approaches are quite orthogonal. Separation logic extends the assertion language with additional operators, while the theory of heap  provides an interchange format for encoding programs with the goal of preserving as much information about the heap as possible. Both could be used in a complementary way to encode program assertions and the program itself. 

\paragraph{Linear Maps} provide a similar proof strategy to that of separation logic, while staying within the confines of classical logic \cite{DBLP:conf/plpv/LahiriQW11}. The authors describe a two-way \emph{erasure transformation}, transforming between imperative programs with a single unified heap and programs with multiple disjoint linear maps. Since the transformation is completely in classical logic, off-the-shelf SMT solvers and theorem provers can be used without a special decision procedure by making use of the existing theories such as the theory of arrays and the theory of sets.

Unlike the transformational approach of linear maps, the theory of heap aims to defer the handling of heap to the solvers. In fact, the linear maps strategy could also make use of the theory of heap in order to have access to more specialised decision procedures, and not be restricted to the theory of arrays. 

\paragraph{Other related work}
The authors of \cite{DBLP:conf/atva/RakamaricBHC07} extend an SMT solver with a decision procedure to decide unbounded heap reachability with support for Boolean and integer data fields. \cite{Lahiri2007ADP} also describes a decision procedure for verification of heap-manipulating programs. Both papers are about verifying heap reachability, and both of them highlight the need for a standard theory of heap as that would have provided a framework for the research and ease the adoption of proposed decision procedures by different solvers. 

\section{Preliminary Experiments and Conclusions}


We have proposed a theory of heap, along with its syntax and
semantics, and discussed possible alternative definitions and
extensions in Section~\ref{sec:extensions}.  The intention is that the
ideas presented here will initiate discussions, and eventually result
in a common interchange language for programs with heap. As a
long-term goal, we would like to include a heap track also at the
CHC-COMP competition.

\iftrue
In order to highlight the feasibility of using the theory in a more concrete setting,
we collected C benchmarks from SV-COMP's \emph{ReachSafety} and 
\emph{MemSafety} categories\footnote{\url{https://github.com/sosy-lab/sv-benchmarks}},
and extended
TriCera\footnote{\url{https://github.com/uuverifiers/tricera/tree/heaptheory}}, 
a model checker for C programs, in order to produce the CHCs in the theory of heap.
To create a preliminary set of CHC benchmarks modulo heap, we filtered
out programs that require heap, but none of the features not yet supported
in our setting (e.g., stack pointers or arrays). In the end, 111 unique benchmarks remained.

To experiment with those benchmarks, the SMT solver
\princess~\cite{princess08} was extended to support the theory using the reasoning and
interpolation procedures from \cite{10.1007/978-3-030-68446-4_9}, and 
the CHC solver \eldarica~\cite{eldarica} was extended to make use of the newly added theory in
\princess. We have made available the benchmarks and the version of \eldarica~used 
during the experiments.\footnote{\url{https://github.com/uuverifiers/eldarica/releases/tag/v2.0.5-heap}}

\smallskip
\hspace*{-\parindent}\begin{minipage}{0.62\textwidth}
The experiments were run on an AMD Opteron 2220 SE machine with 64-bit Linux. The results are given in Table~\ref{table:results}. \eldarica~could solve 26 benchmarks, while others timed out (T/O) after 600 seconds or were unsolvable due to quantified interpolants (as stipulated by Lemma~\ref{lem:interpolation}).
\end{minipage}\hfill
\begin{minipage}{0.33\textwidth}
\centering
\begin{tabular}{rcl|c|c|c}
sat&/&unsat	& t/o	& other & total\\\hline
8&/&18		& 40	& 45	& 111  \\
\end{tabular}
\captionof{table}{Results for \eldarica~2.0.5-heap} \label{table:results}
\end{minipage}
\smallskip

%
\fi

In order to show how the same benchmarks could be encoded using the theory of arrays,
we also provide the array theory versions, which were  translated using
the encoding shown in Listing~\ref{lst-array-encoding}. At the time of writing this
paper, none of the other current CHC solvers that we know of could solve this 
particular encoding of the benchmarks, mostly due to not supporting the theory
combination of ADTs and arrays.

It has to be stressed that the experiments are early, and no
conclusions should be drawn other than that real-world C programs can
indeed be encoded and analysed using the proposed theory.  The algorithms from
\cite{10.1007/978-3-030-68446-4_9} used in the experiments are direct
and unrefined adaptions of procedures for the theory of arrays, and
more work is needed to obtain, e.g., practical interpolation
methods. However, now that the design choice is shifted to the
solvers, alternative approaches can be employed to improve the results
without changing the CHC representation of programs. In this context,
two directions we are currently pursuing are improved decision and
interpolation procedures for the heap theory, and the adaptation of
the invariant-based heap encoding used in JayHorn~\cite{jayhorn-2017}.


\clearpage
\bibliographystyle{splncs04}
\bibliography{refs}

\begin{thebibliography}{10}
\providecommand{\url}[1]{\texttt{#1}}
\providecommand{\urlprefix}{URL }
\providecommand{\doi}[1]{https://doi.org/#1}

\bibitem{cvc4}
Barrett, C., Conway, C.L., Deters, M., Hadarean, L., Jovanovi{'{c}}, D., King,
  T., Reynolds, A., Tinelli, C.: {CVC4}. In: Gopalakrishnan, G., Qadeer, S.
  (eds.) Proceedings of the 23rd International Conference on Computer Aided
  Verification (CAV '11). Lecture Notes in Computer Science, vol.~6806, pp.
  171--177. Springer (Jul 2011),
  \url{http://www.cs.stanford.edu/~barrett/pubs/BCD+11.pdf}, snowbird, Utah

\bibitem{smt-2.6}
Barrett, C., Fontaine, P., Tinelli, C.: {The SMT-LIB Standard: Version 2.6}.
  Tech. rep., Department of Computer Science, The University of Iowa (2017),
  available at {\tt www.SMT-LIB.org}

\bibitem{DBLP:conf/birthday/BjornerGMR15}
Bj{\o}rner, N., Gurfinkel, A., McMillan, K.L., Rybalchenko, A.: Horn clause
  solvers for program verification. In: Beklemishev, L.D., Blass, A.,
  Dershowitz, N., Finkbeiner, B., Schulte, W. (eds.) Fields of Logic and
  Computation {II} - Essays Dedicated to Yuri Gurevich on the Occasion of His
  75th Birthday. Lecture Notes in Computer Science, vol.~9300, pp. 24--51.
  Springer (2015). \doi{10.1007/978-3-319-23534-9\_2},
  \url{https://doi.org/10.1007/978-3-319-23534-9\_2}

\bibitem{DBLP:conf/sas/BjornerMR13}
Bj{\o}rner, N., McMillan, K.L., Rybalchenko, A.: On solving universally
  quantified {Horn} clauses. In: Logozzo, F., F{\"{a}}hndrich, M. (eds.) Static
  Analysis - 20th International Symposium, {SAS} 2013, Seattle, WA, USA, June
  20-22, 2013. Proceedings. Lecture Notes in Computer Science, vol.~7935, pp.
  105--125. Springer (2013). \doi{10.1007/978-3-642-38856-9\_8},
  \url{https://doi.org/10.1007/978-3-642-38856-9\_8}

\bibitem{calculus-of-computation}
Bradley, A.R., Manna, Z.: The calculus of computation - decision procedures
  with applications to verification. Springer (2007).
  \doi{10.1007/978-3-540-74113-8},
  \url{https://doi.org/10.1007/978-3-540-74113-8}

\bibitem{DBLP:journals/fuin/AngelisFPP17a}
{De Angelis}, E., Fioravanti, F., Pettorossi, A., Proietti, M.: Program
  verification using constraint handling rules and array constraint
  generalizations. Fundam. Inform.  \textbf{150}(1),  73--117 (2017).
  \doi{10.3233/FI-2017-1461}, \url{https://doi.org/10.3233/FI-2017-1461}

\bibitem{DBLP:journals/jacm/DowneyS78}
Downey, P.J., Sethi, R.: Assignment commands with array references. J. {ACM}
  \textbf{25}(4),  652--666 (1978). \doi{10.1145/322092.322104},
  \url{https://doi.org/10.1145/322092.322104}

\bibitem{heapTheoryHCVS}
Esen, Z., R{\"{u}}mmer, P.: Towards an {SMT-LIB} theory of heap (extended
  abstract). In: Fribourg, L., Heizmann, M. (eds.) 8th International Workshop
  on Verification and Program Transformation and 7th Workshop on Horn Clauses
  for Verification and Synthesis, VPT/HCVS@ETAPS 2020 2020, and 7th Workshop on
  Horn Clauses for Verification and SynthesisDublin, Ireland, 25-26th April
  2020. {EPTCS}, vol.~320 (2020)

\bibitem{10.1007/978-3-030-68446-4_9}
Esen, Z., R{\"u}mmer, P.: Reasoning in the theory of heap: Satisfiability and
  interpolation. In: Fern{\'a}ndez, M. (ed.) Logic-Based Program Synthesis and
  Transformation. pp. 173--191. LNCS, Springer, Cham (2021)

\bibitem{DBLP:conf/pldi/GrebenshchikovLPR12}
Grebenshchikov, S., Lopes, N.P., Popeea, C., Rybalchenko, A.: Synthesizing
  software verifiers from proof rules. In: Vitek, J., Lin, H., Tip, F. (eds.)
  {ACM} {SIGPLAN} Conference on Programming Language Design and Implementation,
  {PLDI} '12, Beijing, China - June 11 - 16, 2012. pp. 405--416. {ACM} (2012).
  \doi{10.1145/2254064.2254112}, \url{https://doi.org/10.1145/2254064.2254112}

\bibitem{seahorn}
Gurfinkel, A., Kahsai, T., Komuravelli, A., Navas, J.A.: The seahorn
  verification framework. In: Kroening, D., Pasareanu, C.S. (eds.) Computer
  Aided Verification - 27th International Conference, {CAV} 2015, San
  Francisco, CA, USA, July 18-24, 2015, Proceedings, Part {I}. Lecture Notes in
  Computer Science, vol.~9206, pp. 343--361. Springer (2015).
  \doi{10.1007/978-3-319-21690-4\_20},
  \url{https://doi.org/10.1007/978-3-319-21690-4\_20}

\bibitem{DBLP:journals/cacm/Hoare69}
Hoare, C.A.R.: An axiomatic basis for computer programming. Commun. {ACM}
  \textbf{12}(10),  576--580 (1969). \doi{10.1145/363235.363259},
  \url{https://doi.org/10.1145/363235.363259}

\bibitem{eldarica}
Hojjat, H., R{\"{u}}mmer, P.: The {ELDARICA} horn solver. In: Bj{\o}rner, N.,
  Gurfinkel, A. (eds.) 2018 Formal Methods in Computer Aided Design, {FMCAD}
  2018, Austin, TX, USA, October 30 - November 2, 2018. pp.~1--7. {IEEE}
  (2018). \doi{10.23919/FMCAD.2018.8603013},
  \url{https://doi.org/10.23919/FMCAD.2018.8603013}

\bibitem{Iosif2018EncodingSL}
Iosif, R., Serban, C., Reynolds, A., Sighireanu, M.: Encoding separation logic
  in smt-lib v2.5 (2018), \url{https://sl-comp.github.io/docs/smtlib-sl.pdf}

\bibitem{jayhorn-2017}
Kahsai, T., Kersten, R., R{\"{u}}mmer, P., Sch{\"{a}}f, M.: Quantified heap
  invariants for object-oriented programs. In: Eiter, T., Sands, D. (eds.)
  LPAR-21, 21st International Conference on Logic for Programming, Artificial
  Intelligence and Reasoning, Maun, Botswana, May 7-12, 2017. EPiC Series in
  Computing, vol.~46, pp. 368--384. EasyChair (2017),
  \url{https://easychair.org/publications/paper/Pmh}

\bibitem{DBLP:conf/ndss/KalraGDS18}
Kalra, S., Goel, S., Dhawan, M., Sharma, S.: {ZEUS:} analyzing safety of smart
  contracts. In: 25th Annual Network and Distributed System Security Symposium,
  {NDSS} 2018, San Diego, California, USA, February 18-21, 2018. The Internet
  Society (2018),
  \url{http://wp.internetsociety.org/ndss/wp-content/uploads/sites/25/2018/02/ndss2018\_09-1\_Kalra\_paper.pdf}

\bibitem{DBLP:conf/sigsoft/KapurMZ06}
Kapur, D., Majumdar, R., Zarba, C.G.: Interpolation for data structures. In:
  Young, M., Devanbu, P.T. (eds.) Proceedings of the 14th {ACM} {SIGSOFT}
  International Symposium on Foundations of Software Engineering, {FSE} 2006,
  Portland, Oregon, USA, November 5-11, 2006. pp. 105--116. {ACM} (2006).
  \doi{10.1145/1181775.1181789}, \url{https://doi.org/10.1145/1181775.1181789}

\bibitem{DBLP:conf/fmcad/KomuravelliBGM15}
Komuravelli, A., Bj{\o}rner, N., Gurfinkel, A., McMillan, K.L.: Compositional
  verification of procedural programs using {Horn} clauses over integers and
  arrays. In: Kaivola, R., Wahl, T. (eds.) Formal Methods in Computer-Aided
  Design, {FMCAD} 2015, Austin, Texas, USA, September 27-30, 2015. pp. 89--96.
  {IEEE} (2015)

\bibitem{Lahiri2007ADP}
Lahiri, S., Qadeer, S.: A decision procedure for well-founded reachability
  (2007)

\bibitem{DBLP:conf/plpv/LahiriQW11}
Lahiri, S.K., Qadeer, S., Walker, D.: Linear maps. In: Jhala, R., Swierstra, W.
  (eds.) Proceedings of the 5th {ACM} Workshop Programming Languages meets
  Program Verification, {PLPV} 2011, Austin, TX, USA, January 29, 2011. pp.
  3--14. {ACM} (2011). \doi{10.1145/1929529.1929531},
  \url{https://doi.org/10.1145/1929529.1929531}

\bibitem{boogieTypeEncoding2010}
Leino, K.R.M., R{\"u}mmer, P.: A polymorphic intermediate verification
  language: Design and logical encoding. In: Esparza, J., Majumdar, R. (eds.)
  Tools and Algorithms for the Construction and Analysis of Systems. LNCS,
  vol.~6015, pp. 312--327. Springer (2010)

\bibitem{rusthorn}
Matsushita, Y., Tsukada, T., Kobayashi, N.: Rusthorn: Chc-based verification
  for rust programs. In: M{\"{u}}ller, P. (ed.) Programming Languages and
  Systems - 29th European Symposium on Programming, {ESOP} 2020, Held as Part
  of the European Joint Conferences on Theory and Practice of Software, {ETAPS}
  2020, Dublin, Ireland, April 25-30, 2020, Proceedings. Lecture Notes in
  Computer Science, vol. 12075, pp. 484--514. Springer (2020).
  \doi{10.1007/978-3-030-44914-8\_18},
  \url{https://doi.org/10.1007/978-3-030-44914-8\_18}

\bibitem{DBLP:conf/ifip/McCarthy62}
McCarthy, J.: Towards a mathematical science of computation. In: Information
  Processing, Proceedings of the 2nd {IFIP} Congress 1962, Munich, Germany,
  August 27 - September 1, 1962. pp. 21--28. North-Holland (1962)

\bibitem{DBLP:conf/tacas/McMillan04}
McMillan, K.L.: An interpolating theorem prover. In: Jensen, K., Podelski, A.
  (eds.) Tools and Algorithms for the Construction and Analysis of Systems,
  10th International Conference, {TACAS} 2004, Held as Part of the Joint
  European Conferences on Theory and Practice of Software, {ETAPS} 2004,
  Barcelona, Spain, March 29 - April 2, 2004, Proceedings. Lecture Notes in
  Computer Science, vol.~2988, pp. 16--30. Springer (2004).
  \doi{10.1007/978-3-540-24730-2\_2},
  \url{https://doi.org/10.1007/978-3-540-24730-2\_2}

\bibitem{DBLP:conf/sas/MonniauxG16}
Monniaux, D., Gonnord, L.: Cell morphing: From array programs to array-free
  {Horn} clauses. In: Rival, X. (ed.) Static Analysis - 23rd International
  Symposium, {SAS} 2016, Edinburgh, UK, September 8-10, 2016, Proceedings.
  Lecture Notes in Computer Science, vol.~9837, pp. 361--382. Springer (2016).
  \doi{10.1007/978-3-662-53413-7\_18},
  \url{https://doi.org/10.1007/978-3-662-53413-7\_18}

\bibitem{z3}
de~Moura, L.M., Bj{\o}rner, N.: {Z3:} an efficient {SMT} solver. In:
  Ramakrishnan, C.R., Rehof, J. (eds.) Tools and Algorithms for the
  Construction and Analysis of Systems, 14th International Conference, {TACAS}
  2008, Held as Part of the Joint European Conferences on Theory and Practice
  of Software, {ETAPS} 2008, Budapest, Hungary, March 29-April 6, 2008.
  Proceedings. Lecture Notes in Computer Science, vol.~4963, pp. 337--340.
  Springer (2008). \doi{10.1007/978-3-540-78800-3\_24},
  \url{https://doi.org/10.1007/978-3-540-78800-3\_24}

\bibitem{DBLP:conf/aplas/PerezR13}
P{\'{e}}rez, J.A.N., Rybalchenko, A.: Separation logic modulo theories. In:
  Shan, C. (ed.) Programming Languages and Systems - 11th Asian Symposium,
  {APLAS} 2013, Melbourne, VIC, Australia, December 9-11, 2013. Proceedings.
  Lecture Notes in Computer Science, vol.~8301, pp. 90--106. Springer (2013).
  \doi{10.1007/978-3-319-03542-0\_7},
  \url{https://doi.org/10.1007/978-3-319-03542-0\_7}

\bibitem{DBLP:conf/atva/RakamaricBHC07}
Rakamaric, Z., Bruttomesso, R., Hu, A.J., Cimatti, A.: Verifying
  heap-manipulating programs in an {SMT} framework. In: Namjoshi, K.S., Yoneda,
  T., Higashino, T., Okamura, Y. (eds.) Automated Technology for Verification
  and Analysis, 5th International Symposium, {ATVA} 2007, Tokyo, Japan, October
  22-25, 2007, Proceedings. Lecture Notes in Computer Science, vol.~4762, pp.
  237--252. Springer (2007). \doi{10.1007/978-3-540-75596-8\_18},
  \url{https://doi.org/10.1007/978-3-540-75596-8\_18}

\bibitem{DBLP:conf/atva/ReynoldsIS016}
Reynolds, A., Iosif, R., Serban, C., King, T.: A decision procedure for
  separation logic in {SMT}. In: Artho, C., Legay, A., Peled, D. (eds.)
  Automated Technology for Verification and Analysis - 14th International
  Symposium, {ATVA} 2016, Chiba, Japan, October 17-20, 2016, Proceedings.
  Lecture Notes in Computer Science, vol.~9938, pp. 244--261 (2016).
  \doi{10.1007/978-3-319-46520-3\_16},
  \url{https://doi.org/10.1007/978-3-319-46520-3\_16}

\bibitem{DBLP:conf/lics/Reynolds02}
Reynolds, J.C.: Separation logic: {A} logic for shared mutable data structures.
  In: 17th {IEEE} Symposium on Logic in Computer Science {(LICS} 2002), 22-25
  July 2002, Copenhagen, Denmark, Proceedings. pp. 55--74. {IEEE} Computer
  Society (2002). \doi{10.1109/LICS.2002.1029817},
  \url{https://doi.org/10.1109/LICS.2002.1029817}

\bibitem{DBLP:conf/pldi/RondonKJ08}
Rondon, P.M., Kawaguchi, M., Jhala, R.: Liquid types. In: Gupta, R.,
  Amarasinghe, S.P. (eds.) Proceedings of the {ACM} {SIGPLAN} 2008 Conference
  on Programming Language Design and Implementation, Tucson, AZ, USA, June
  7-13, 2008. pp. 159--169. {ACM} (2008). \doi{10.1145/1375581.1375602},
  \url{https://doi.org/10.1145/1375581.1375602}

\bibitem{princess08}
R{\"u}mmer, P.: A constraint sequent calculus for first-order logic with linear
  integer arithmetic. In: Proceedings, 15th International Conference on Logic
  for Programming, Artificial Intelligence and Reasoning. LNCS, vol.~5330, pp.
  274--289. Springer (2008)

\bibitem{DBLP:journals/corr/abs-2008-02939}
R{\"{u}}mmer, P.: Competition report: {CHC-COMP-20}. In: Fribourg, L.,
  Heizmann, M. (eds.) Proceedings 8th International Workshop on Verification
  and Program Transformation and 7th Workshop on Horn Clauses for Verification
  and Synthesis, VPT/HCVS@ETAPS 2020 2020, and 7th Workshop on Horn Clauses for
  Verification and SynthesisDublin, Ireland, 25-26th April 2020. {EPTCS},
  vol.~320, pp. 197--219 (2020). \doi{10.4204/EPTCS.320.15},
  \url{https://doi.org/10.4204/EPTCS.320.15}

\bibitem{DBLP:conf/pepm/SatoI019}
Sato, R., Iwayama, N., Kobayashi, N.: Combining higher-order model checking
  with refinement type inference. In: Hermenegildo, M.V., Igarashi, A. (eds.)
  Proceedings of the 2019 {ACM} {SIGPLAN} Workshop on Partial Evaluation and
  Program Manipulation, PEPM@POPL 2019, Cascais, Portugal, January 14-15, 2019.
  pp. 47--53. {ACM} (2019). \doi{10.1145/3294032.3294081},
  \url{https://doi.org/10.1145/3294032.3294081}

\end{thebibliography}

	

\appendix
\newpage

\section{Proof of Lemma~\ref{lem:NP}}

  \emph{Membership in NP} follows from the polynomial-time reduction
  of heap formulas to array formulas
  (Section~\ref{sec:encoding-array}), and NP-completeness of the
  theory of arrays combined with NP theories to represent
  indexes \cite{calculus-of-computation}.

  \emph{NP-hardness} follows using a similar reduction of the Boolean
  SAT problem as in \cite{DBLP:journals/jacm/DowneyS78}. Suppose
  $C_1 \wedge \cdots \wedge C_k$ is a conjunction of clauses over
  Boolean variables~$x_1, \ldots, x_m$. We introduce object
  variables~$T, F : O$ to represent truth values, and for each Boolean
  variable~$x_i$ two address variables~$a_i, \bar a_i$. We then create
  a heap~$h$ with exactly two valid addresses, and represent each clause
  as a chain of $\wt$ operations. The resulting heap formula is
  equisatisfiable to $C_1 \wedge \cdots \wedge C_k$:
  \begin{align*}
    & T \not= F \wedge h = \alloc(\alloc(\eh, F).\_1, F).\_1
    \\
     \wedge~~ & \bigwedge_{i=1}^m
      \big(\ia(h, a_i) \wedge \ia(h, \bar a_i) \wedge a_i \not= \bar a_i \big)
    \\
    \wedge~~ & \bigwedge_{i=1}^k
               T = \rd(W_i, \ntha_1)
  \end{align*}
  where the term~$W_i$ for a clause~$C_i$ is defined by:
  \begin{align*}
    W_i &= \wt(\cdots\wt(\wt(h, t^i_1, T), t^i_2, T), \cdots, t^i_m, T)
    \\
    t^i_j &=
          \begin{cases}
            \ntha_1 & \text{if both~} x_j \text{~and~} \neg x_j \text{~occur in~} C_i
            \\
            a_j & \text{if~} x_j \text{~occurs in~} C_i
            \\
            \bar a_j & \text{if~} \neg x_j \text{~occurs in~} C_i
            \\
            \nullAddr & \text{otherwise}
          \end{cases}
  \end{align*}

\section{Proof of Lemma~\ref{lem:interpolation}}

  This result carries over from the result for
  arrays~\cite{DBLP:conf/sigsoft/KapurMZ06,DBLP:conf/tacas/McMillan04}. As
  an example, consider the following for $A$ and $B$:
\begin{align*}
A:&~h_2 = \wt(h_1,p_1,o_1) \wedge \ia(h_1,p_1)\\
B:&~p_2 \neq p_3 \wedge \rd(h_2,p_2) \neq \rd(h_1,p_2) \wedge \rd(h_2,p_3) \neq \rd(h_1,p_3)\\&~\wedge \ia(h_1,p_2) \wedge \ia(h_1,p_3)
\end{align*}
Then the only possible interpolants are quantified ones such as:
\begin{align*}
\forall{x,y.}&~(x=y \vee \rd(h_1,x) = \rd(h_2,x) \vee 
        \rd(h_1,y) = \rd(h_2,y)~\vee \\
        &~\neg \ia(h_1,x) \vee \neg \ia(h_1,y))
\end{align*}

\end{document}